\documentclass[twocolumn,showpacs,preprintnumbers,amsmath,amssymb,nofootinbib]{revtex4}
\usepackage{graphicx}
\usepackage{dcolumn}
\usepackage{bm}
\usepackage{amsmath}    
\usepackage{graphicx}   

\newcommand{\be}{\begin{equation}}
\newcommand{\ee}{\end{equation}}
\newcommand{\bea}{\begin{eqnarray}}
\newcommand{\eea}{\end{eqnarray}}

\newcommand{\ve}{\varepsilon}

\begin{document}

\title{Pion-nucleon correlations in finite nuclei in a relativistic framework: effects on the shell structure}

\author{Elena Litvinova}
\affiliation{Department of Physics, Western Michigan University,
Kalamazoo, MI 49008-5252, USA } \affiliation{National
Superconducting Cyclotron Laboratory, Michigan State University,
East Lansing, MI 48824-1321, USA}
\date{\today}
\begin{abstract}
The relativistic particle-vibration coupling (RPVC) model  is extended by the inclusion of 
isospin-flip
excitation modes into the phonon space, introducing a new mechanism of dynamical interaction between nucleons with different isospin in the nuclear medium.  Protons and neutrons exchange by collective modes which are formed by isovector $\pi$ and $\rho$-mesons, in turn, softened considerably because of coupling to nucleons of the medium. These modes are investigated within the proton-neutron relativistic  random phase approximation (pn-RRPA) and relativistic proton-neutron time blocking approximation (pn-RTBA). The appearance of  isospin-flip states with sizable transition probabilities at low energies points out that they are likely to couple to the single-particle degrees of freedom and, in addition to isoscalar low-lying phonons, to modify their spectroscopic characteristics. Such a coupling is quantified for the shell structure of $^{100,132}$Sn and found significant for the location of the dominant single-particle states. 

 \end{abstract}


\maketitle

\section{Introduction}

A consistent treatment of pion degrees of freedom is still one of the challenges for nuclear structure theory, despite the fact that the meson-exchange nucleon-nucleon interaction is known since the work of Yukawa \cite{Yukawa}. Being a mediator of the medium-range nucleon-nucleon interaction, the pion played the central role in approaches 
developed by the so-called 'Moscow school' \cite{Mig.74,TSMM.75,STF.75,MSMM.76,FST.77,TKC.77,KM.77,FST.79,BSTF.81,MSTV.90} and 'Stony Brook school' \cite{BB.73,BW.76,BCDM.75,RGG.76,TW.78,TW.79,Meyer.81,DFMW.83}. Characteristics of the pion bound in nuclear medium were extensively discussed, in particular, in the context of possible formation of pion condensate in finite nuclei and various phases of nuclear matter \cite{Meyer.81,DFMW.83,MSTV.90}. Later on, chiral effective field theory (EFT) based on the approximate spontaneously broken chiral symmetry of QCD emerged from the pioneering work of Weinberg \cite{Weinberg}. Since then, ab initio methods based on chiral EFT demonstrate an increasing success in the description of nuclear structure properties although their applicability is still limited to light nuclei only \cite{EHM.09,ME.11}. 

Many-body methods for sufficiently heavy nuclear systems (A$>$16) include various nucleon-nucleon correlations on top of a mean field, which serves as a redefined vacuum constructed as an approximate realization of Kohn-Shahm density functional theory. The original (non-relativistic) nuclear field theory (NFT) is based on the idea of Bohr and Mottelson \cite{BM.75} about coupling between single-particle and vibrational degrees of freedom in atomic nuclei, often referred to as particle-vibration coupling (PVC). This idea has been developed over the years \cite{H.69,RW.73,BBDLP.76,BBBL.77}
 and implemented very successfully in 
 approaches known as a non-relativistic NFT 
 \cite{BBB.83,BBBD.85,BB.81,CBGBB.92,CB.01,SBC.04,GRB.05,IBV.12,IPBVB.15,PBVB.10,PBMIVB.11}, extensions of the Landau-Migdal theory \cite{KTT.97,KST.04,LT.07,TSG.07} or quasiparticle-phonon model \cite{SSV.82,GSV.88,Sol.92}. Lately, self-consistent versions of the PVC model have become available in both non-relativistic \cite{NCV.14,CCSB.14} and relativistic \cite{LR.06,LA.11,L.12,LRT.07,LRT.08,LRT.10,LRT.13,L1.15,MLVR.12,LBFMZ.14} formulations. In the majority of the non-relativistic self-consistent calculations the PVC effects are studied in doubly-magic nuclei, where superfluidity effects do not appear, which simplifies calculations considerably. Studies of  these effects in open-shell nuclei have become possible relatively recently in a non-relativistic approach for single-quasiparticle states \cite{IBV.12,IPBVB.15} and and in a self-consistent relativistic approach for both single-quasiparticle spectra \cite{L.12} and response \cite{LRT.08}. 
 
The  covariant (relativistic) version of the 
nuclear
field theory \cite{LR.06,LA.11,L.12,LRT.07,LRT.08,LRT.10,LRT.13,L1.15,MLVR.12,LBFMZ.14,L2.15} is based on the quantum hadrodynamics (QHD) Lagrangian, which, in contrast to chiral EFT, includes heavy mesons explicitly, although their masses and coupling constants are adjusted to bulk properties of medium-mass nuclei (see review \cite{VALR.05} and references therein).  Numerous calculations within these models have shown that the effects of quasiparticle-vibration coupling (QVC) modify nuclear shell structure and spectra of excitations considerably and bring the results in essentially better agreement to experimental data, in some cases approaching spectroscopic accuracy \cite{LRT.08,E.10,M.12,L.12,LRT.13}. In particular, it turned out that the effects of QVC in open-shell nuclei are quantitatively stronger than those in closed-shell ones, which can be attributed to the existence of phonons with very low energies ($\sim$1 MeV or even lower) in medium-mass open-shell nuclei, compared to $\sim$4 MeV in medium-mass doubly-magic systems. Obviously, superfluidity plays a very important role providing better conditions for the QVC, which can be seen from the structure of the QVC self-energy \cite{L.12}. The pion field does not contribute on the mean-field level, if the parity conservation is imposed, however, its dynamical contribution is decisive for the description of spin-isospin transitions \cite{PNVR.04,NMVPR.05,MLVR.12,LBFMZ.14}. 

Recently, it has
been realized that the spin-isospin interaction, especially the tensor interaction, is of particular and direct importance for single-particle properties of exotic nuclei \cite{Otsuka,BFN.08}.  This type of interaction is commonly associated  with one-pion exchange, although other mesons can also give contributions of the tensor type, which has been obtained in the relativistic Hartree-Fock \cite{Long} and relativistic RPA (RRPA) calculations for spin-isospin-flip transitions \cite{Liang}. However, the latter calculations are confined by the static approximation for the pion field, neglecting retardation effects, and include only the first-order contribution  to the pion-exchange interaction. Nevertheless, RPA in the spin-isospin channel gives a sizable softening of the pion modes, where the softness is caused by medium polarization effects, as it has been recognized already in earlier calculations  \cite{MSTV.90}. The appearance of such modes at low energies was interpreted as a signature for the onset of the pion condensation, although it was not clear which observables are sensitive enough to the presence of pion condensate and which of those can be detected experimentally with reasonable precision.

The main goal of the present work is to reveal dynamical contributions of the pion-exchange interaction to nuclear structure properties. The most important dynamical contributions are identified as terms in the nucleonic self-energy, which represent exchange by collective modes. Technically, besides the isoscalar modes, which are exchanged between nucleons of the same isospin and commonly considered in NFT,  the isovector modes exchanged between nucleons with different isospin are added. 
First of all, the approach  of Refs. \cite{PNVR.04,MLVR.12}, namely the proton-neutron relativistic random phase approximation (pn-RRPA) and proton-neutron relativistic time blocking approximation (pn-RTBA) is applied to calculations of the isospin-flip and spin-isospin-flip excitations. The obtained spectra are evaluated to identify the low-lying isovector modes of excitation, which give the most important contribution to the nucleonic self-energy, according to their degree of collectivity. Then, the selected most collective modes are coupled to the single-particle degrees of freedom by means of introducing corresponding terms into the single-nucleon self-energy. 
Thus, on one hand, the relativistic NFT is extended by isospin vibrations and, on the other hand, dynamical effects of the pion exchange are introduced into the theory beyond the Hartree-Fock approximation. The effects of these new terms on the single-particle characteristics, such as their energies and spectroscopic factors, are discussed.  In the context of the discussion of pion condensate, it is shown that nuclear single-particle characteristics can represent observables which are sensitive  to the presence of soft pion modes.  

\section{Theoretical framework}

\subsection{Nucleonic self-energy and Dyson's equation }

The motion of protons and neutrons in nuclei is quantified by
one-nucleon self-energy, or mass operator. In contrast to free nucleons, protons and neutrons in the nuclear medium are exposed to strong polarization effects and, hence, their motion is modified considerably. In the simplest mean field (Hartree or Hartree-Fock) approximation,  this effect can be described by introducing an energy-independent effective mass $m^{\ast}$ instead of the bare nucleon mass, which helps to reproduce bulk nuclear properties reasonably well, but gives a very poor description of the single-particle spectra.
The latter problem can only be solved by going beyond the mean-field approximation and by introducing an energy-dependent effective mass.
The most important origin of the energy dependence is given by the coupling of the single particle motion to low-lying collective vibrations \cite{H.69,RW.73,BBDLP.76,BBBL.77}, see also the discussion in Ref. \cite{LR.06}, where a relativistic particle-vibration coupling (RPVC) model was originally proposed.
In both the RPVC model and relativistic quasiparticle-vibration coupling (RQVC) model for systems with pairing correlations \cite{L.12}, the energy-independent part of the nucleonic self-energy is given by the relativistic mean field (relativistic Hartree (RH) or Hartree-Bogoliubov (RHB)) approximation. The RH/RHB eigenstates $\{k,\eta\}$  are used as a basis for calculating the dynamical part of the self-energy, matrix elements of which in this basis read:
\bea
\Sigma^{(e)\eta_1\eta_2}_{k_1k_2}(\ve) = \sum\limits_{\eta_{\mu}=\pm1}\sum\limits_{\eta=\pm1}\sum\limits_{k,\mu} \frac{\delta_{\eta\eta_{\mu}}\gamma^{\eta_{\mu};\eta_1\eta}_{\mu;k_1k}
\gamma^{\eta_{\mu};\eta_2\eta\ast}_{\mu;k_2k}}{\ve - \eta E_k - \eta_{\mu}(\Omega_{\mu} - i\delta)}, \label{se}\\
\delta \to +0,\ \ \ \  \ \ \ \ \  \nonumber
\eea
where the indices $k, k_1, k_2$ represent the complete set of the single-particle quantum numbers, and the indices $\eta, \eta_1, \eta_2$ indicate the upper and lower components of the matrix elements in the Nambu space. In the limit of no superfluidity $\eta_1 = 1$, if the state $k_1$ is above the Fermi energy, and $\eta_1= -1$, if the state $k_1$ is below the Fermi energy.  
The index $\mu$ in Eq. (\ref{se}) runs over the set of phonons taken
into account. Their vertices $\gamma_{\mu}$ and frequencies $\Omega_{\mu}$ can be calculated in the self-consistent relativistic quasiparticle random phase approximation, as described in Ref. \cite{LRT.08}, and  $E_k$ are the energies of the Bogoliubov's quasiparticles obtained in the RHB approach, so that the product $\eta E_k$ takes the correct limit of the mean-field energy in the case of no superfluidity.
The index $\eta_{\mu}$ labels forward and backward going diagrams in the self-energy (\ref{se}), see Eq. (\ref{gammazeta}) below.
Using the same representation, the Dyson's equation for the single-quasiparticle Green's function can be formulated as follows:
\bea
\sum\limits_{\eta=\pm 1;k}  \Bigl( (\ve - \eta_1E_{k_1})\delta_{\eta_1\eta}\delta_{k_1k} - \Sigma^{(e)\eta_1\eta}_{k_1k}(\ve)\Bigr)
G^{\eta\eta_2}_{kk_2}(\ve) = \nonumber \\
 \delta_{\eta_1\eta_2}\delta_{k_1k_2}.
 \label{dyson}
\eea
By solving this equation, as described in detail in Ref. \cite{LR.06}, one can obtain the energies of the fragmented states $E^{(\nu)}_k$ and the corresponding spectroscopic factors $S^{(\nu)}_k$, which characterize the probability of occupying the levels $E^{(\nu)}_k$ by quasiparticles.

In the following we will focus on the isospin structure of the self-energy of Eq. (\ref{se}). It represents the lowest-order phonon polarization correction to the single-particle propagator and contains already an infinite sum of the diagrams dressing the fermionic line.  If the summation is performed for the ring type of diagrams, the vertices $\gamma_{\mu}$ and frequencies $\Omega_{\mu}$ are approximated by an R(Q)RPA phonon solution. Formally, the sum in Eq. (\ref{se}) runs over all possible intermediate nucleonic states  $k$. The common practice is to include only the states $k$ of the same isospin as the one of the external states $k_1, k_2$. In the relativistic framework, this practice leads to a very good description of the single-quasiparticle spectra in open-shell nuclei, compared to data, and to a considerable improvement of the results in the closed shell nuclei, although in the latter case the description of the energies have not yet reached spectroscopic accuracy. In this context, the inclusion of components in the self-energy connecting states of different isospin can serve as the first step of the search for possible missing mechanisms of PVC.

The next subsections are devoted to the proton-neutron phonons in nuclei and to establishing their characteristics.

\subsection{Isospin-flip modes in $^{100,132}$Sn}

Isospin-flip modes of excitation in nuclei are the most sensitive probes to reveal the properties of the isospin part of nucleon-nucleon interaction.  Besides that, precise knowledge about various strength distributions involving isospin transfer, especially in exotic nuclei, is of great importance for astrophysics as well as for numerous applications. 

Another aspect of isospin dynamics is its direct connection to pionic degrees of freedom, that can be revealed in calculations which include pion-nucleon interaction explicitly. Already in earlier works (see Ref. \cite{MSTV.90} and references therein) it has been realized that the explicit separation of pion-exchange processes lead to a considerable improvement of the description of isospin transfer modes, beta-decay half-lives and magnetic momenta.

Lately, self-consistent calculations in the relativistic framework have become possible on the (Q)RPA \cite{PNVR.04,NMVPR.05,Liang} and beyond QRPA \cite{MLVR.12,LBFMZ.14} levels, using the free-space pseudovector coupling strength for pion-nucleon interaction.  
In particular, a very reasonable description of the position of the Gamow-Teller resonance (GTR) and beta-decay half-lives in medium-mass nuclei has been achieved already on the QRPA level without a non-relativistic reduction of the pion-exchange interaction \cite{PNVR.04,NMVPR.05}.
Besides that, it has been shown that the explicit inclusion of the meson exchange (Fock) term in both the mean-field and effective interaction gives a relatively good description of the GTR's centroid on the level of self-consistent RPA 
 \cite{Liang}. Calculations beyond R(Q)RPA demonstrate an important role of the phonon coupling for spreading of the  GTR \cite{LBFMZ.14} and spin-dipole resonance (SDR) \cite{MLVR.12}. Analogously to the situation for giant resonances in the neutral channel, the inclusion of the phonon coupling mechanism shows a considerable improvement of the description for both GTR and SDR.   
  
SDR occurs as a nuclear response to the spin-isospin flip operator with an additional one unit angular momentum transfer, i.e. to the operator:
\be
P_{L\pm}^{\lambda} = \sum\limits_{i=1}^{A} r^L(i)[\sigma (i)\otimes Y_L(i)]^{\lambda}\tau_{\pm}(i)
\label{extfield}
\ee
with $L = 1, \lambda = 0,1,2$ and parity $\pi = -1$. Components of different values of $\lambda$ can be disentangled in charge-exchange reaction experiments on stable nuclei and show a hierarchy of the $\lambda = 0,1,2$ centroids from higher to lower energy: $E(2^-) < E(1^-) <  E(0^-)$ \cite{Wakasa}. Although in the calculations based on the relativistic Hartree approach (without Fock term in the mean field) this hierarchy is not always reproduced giving a higher value for the $E(1-)$, the overall SDR distribution agrees with the data very reasonably because of the dominance of the $2^-$ component \cite{MLVR.12}. 
\begin{figure}
\begin{center}
\includegraphics[scale=0.45]{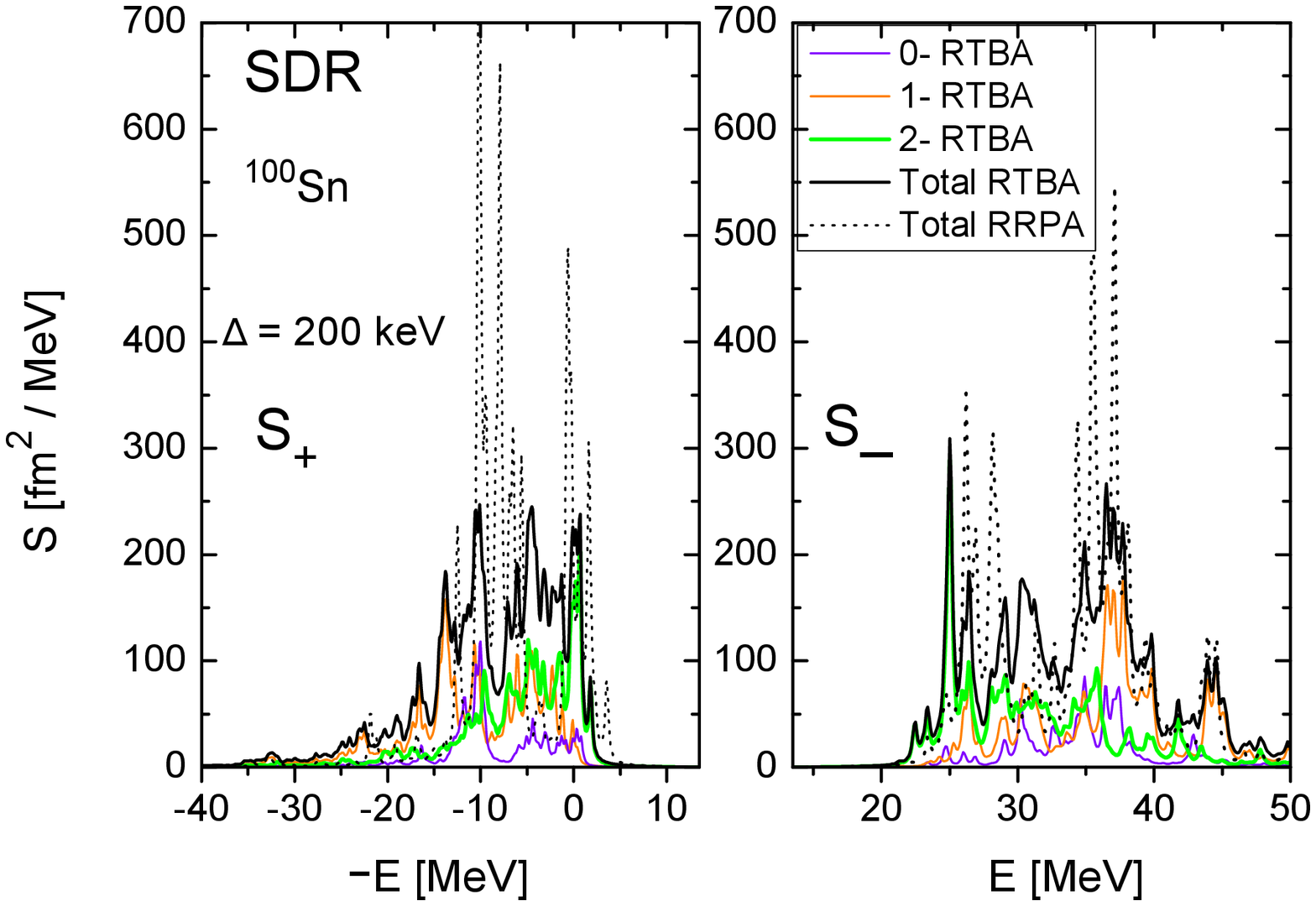}
\includegraphics[scale=0.45]{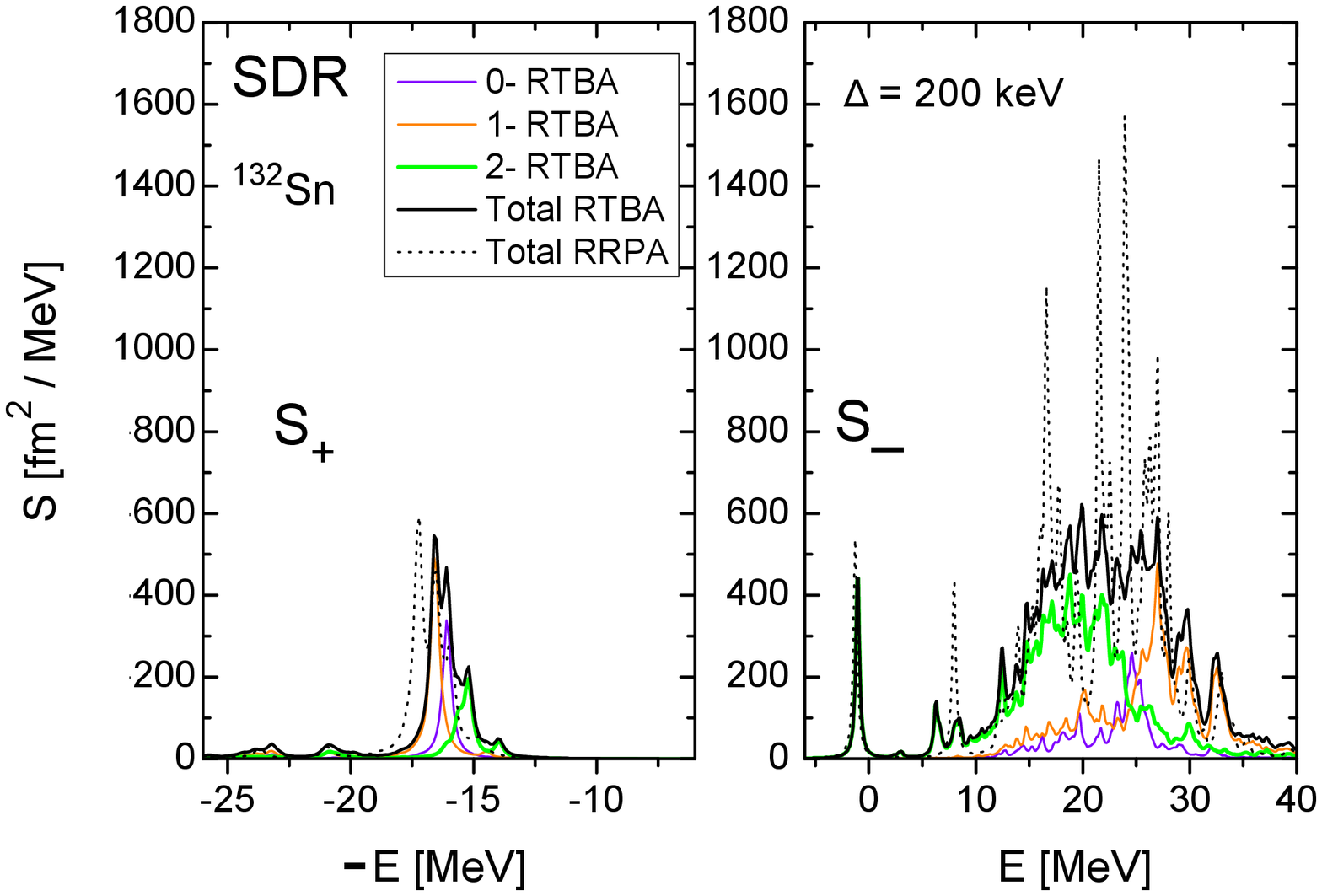}
\end{center}
\caption{Spin dipole resonance in $^{100}$Sn (upper panels) and $^{132}$Sn (lower panels).  The energy scales are relative to Sn nuclei (see text for details).}
\label{fig1}%
\end{figure}
Recently, the latter model has been applied to calculations of the SDR in neutron-rich nuclei, such as $^{132}$Sn and $^{78}$Ni \cite{L2.15}. It has been found that in these nuclei in the $2^-$ component of the SDR a low-energy structure is developed, resembling the well-known low-lying collective quadrupole excitation in the isoscalar channel. Here we consider the case of $^{132}$Sn in more detail and in comparison to a neutron-deficient $^{100}$Sn. Figure \ref{fig1} displays both the $S_-$ and $S_+$ branches of the SDR in the latter nuclei. The $0^-,1^-, 2^-$ components calculated within the pn-RTBA approach of Ref. \cite{MLVR.12} are represented by the violet, orange and green curves, respectively. The black solid curve shows the sum of all  three components and the dashed curve shows the total SDR strength calculated in pn-RRPA, where the particle-vibration coupling mechanism is absent. Thus, the difference between the solid and dashed black curves shows the effect of the SDR fragmentation due to the PVC. Another general observation is that in $^{100}$Sn both branches are well developed and, although not exactly symmetric because of the presence of the Coulomb interaction, exhibit similar strength distributions, while for $^{132}$Sn very little strength is found in the $S_+$ branch, as compared to $S_-$ because of the large isospin asymmetry in this nucleus. 
 
Although the PVC effect is quite noticeable for the overall picture, we are interested here in the properties of the low-lying isospin-flip modes. One can see that in both $^{100}$Sn and $^{132}$Sn strong $2^-$ states form on the low-energy shoulders of the $S_-$ strength distributions. They can bee seen already in the pn-RRPA approach and they are not affected very much by the PVC. In $^{100}$Sn such a state can be seen at about 25 MeV in the $S_-$ and also around 0 MeV in the $S_+$ branch. In $^{132}$Sn such a state in $S_-$ is located at -1.7 MeV and it is visibly separated from the giant resonance. Notice that all the energies are given with respect to the ground state of the $^{100,132}$Sn nuclei and, of course, become positive with respect to the daughter nuclei after the corresponding shifts. Here and in the following we keep the energy scale relative to the ground states of the tin nuclei, in particular, in the self-energies considered below in Eqs. (\ref{se1a}), (\ref{se1b}).

Thus, there appears another aspect of the isospin transfer modes which has been barely discussed in the literature. As low-energy collective excitations can appear in the isovector channel, similarly to the ones in the isoscalar channel, they are likely to couple to single-particle motion. Thus, such modes should be included in the nucleonic self-energy and their contributions should be investigated numerically \cite{IBV.12}. The next sections are devoted to the implementation of this task.

\subsection{Nucleonic self-energy with pion exchange}

As in Refs. \cite{LR.06,L.12} and many other applications, here we use the diagonal approximation for the Eqs. (\ref{se}), (\ref{dyson}).
In a spherical non-superfluid system, the proper neutron and proton self-energies $\Sigma^{(e)}_{n}(\ve)$ and $\Sigma^{(e)}_{p}(\ve)$ including contributions from both isoscalar and isovector phonons have the following form:
\bea
\Sigma^{(e)}_{n}(\ve) = \frac{1}{2j_{n}+1}\sum\limits_{(\mu n^{\prime})} \frac{|\gamma_{(\mu;nn^{\prime})}^{\eta_{n\prime};\eta_n\eta_{n\prime}} |^2}{\ve - \ve_{n\prime} - \eta_{n\prime}(\Omega_{\mu} - i\delta)} + \nonumber \\
\frac{1}{2j_{n}+1}\sum\limits_{(\lambda p^{\prime})} \frac{|\zeta_{(\lambda;np^{\prime})}^{\eta_{p\prime};\eta_n\eta_{p\prime}} |^2}{\ve - \ve_{p\prime} - \eta_{p\prime}(\omega_{\lambda} - i\delta)}, \label{se1a}\\
\Sigma^{(e)}_{p}(\ve) = \frac{1}{2j_{p}+1}\sum\limits_{(\mu p^{\prime})} \frac{|\gamma_{(\mu;pp^{\prime})}^{\eta_{p\prime};\eta_p\eta_{p\prime}}  |^2}{\ve - \ve_{p\prime} - \eta_{p\prime}(\Omega_{\mu} - i\delta)} + \nonumber \\
\frac{1}{2j_{p}+1}\sum\limits_{(\lambda n^{\prime})} \frac{| \zeta_{(\lambda;pn^{\prime})}^{\eta_{n\prime};\eta_p\eta_{n\prime}}  |^2}{\ve - \ve_{n\prime} - \eta_{n\prime}(\omega_{\lambda} - i\delta)}, 
\label{se1b}
\eea
where 
\bea
\gamma_{(\mu;nn^{\prime})}^{\eta_{\mu};\eta_n\eta_{n\prime}} =   \gamma_{(\mu;nn^{\prime})}^{\eta_n\eta_{n^{\prime}}}  \delta_{\eta_{\mu},+1} + \gamma_{(\mu;n^{\prime} n)}^{\eta_{n^{\prime}}\eta_n}  \delta_{\eta_{\mu},-1} \nonumber \\
\zeta_{(\lambda;np^{\prime})}^{\eta_{\lambda};\eta_n\eta_{p\prime}} =  \zeta_{(\lambda;np^{\prime})}^{\eta_n\eta_{p^{\prime}}} \delta_{\eta_{\lambda},+1} + \zeta_{(\lambda;p^{\prime} n)}^{\eta_{p^{\prime}}\eta_n} \delta_{\eta_{\lambda},-1},
\label{gammazeta}
\eea
and
\bea
\gamma_{(\mu;nn^{\prime})}^{\eta_n\eta_{n^{\prime}}}  = \langle n \parallel \gamma_{(\mu)}^{\eta_n\eta_{n^{\prime}}} \parallel n^{\prime}\rangle \nonumber \\
\zeta_{(\lambda;np^{\prime})}^{\eta_n\eta_{p^{\prime}}}  = \langle n \parallel \zeta_{(\lambda)}^{\eta_n\eta_{p^{\prime}}} \parallel p^{\prime}\rangle,
\label{gammazeta}
\eea
where the round brackets indicate the full sets of quantum numbers with excluded magnetic ones (reduced matrix elements), and the analogous convention for the matrix elements with interchanged proton and neutron indices applies. In Eqs. (\ref{se1a}), (\ref{se1b}) we distinguish between the isoscalar phonons with the vertices $\gamma_{\mu}$ and frequencies $\Omega_{\mu}$ and isovector (proton-neutron (pn) and neutron-proton (np)) ones with the vertices $\zeta_{\lambda}$ and frequencies $\omega_{\lambda}$.  
\begin{figure*}
\begin{center}
\vspace{-6cm}
\includegraphics[scale=0.7]{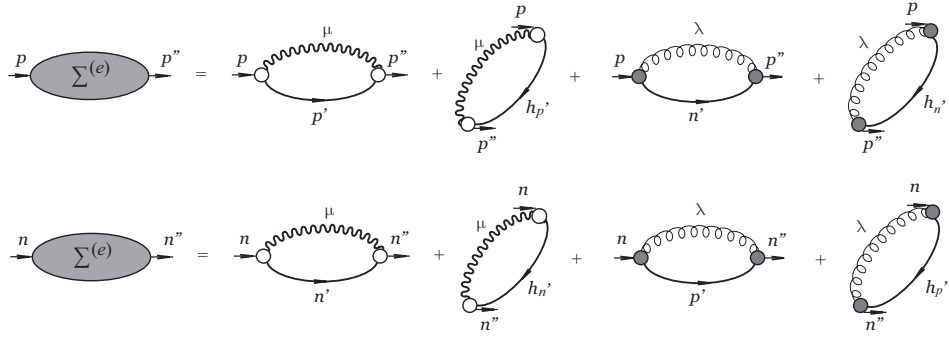}
\end{center}
\caption{Isospin structure of the proper self-energy in the second-order of PVC.  }
\label{figse}%
\end{figure*}
These quantities are to be extracted from the response function of various multipolarities $J_{\mu}$, $J_{\lambda}$ and parities. For the isoscalar phonons this procedure is described in detail in Refs. \cite{LR.06,LRT.08} and consists of the RRPA calculations for the particle-hole (ph) and hole-particle (hp) components of the vertices $\gamma_{\mu}$ and subsequent determination of the particle-particle (pp) and hole-hole (hh) components entering the first sum of the Eqs. (\ref{se1a}), (\ref{se1b}). The characteristics of the isovector phonons are calculated analogously. First, the proton-neutron RRPA (pn-RRPA) equations are solved for the ph-components of the transition densities ${\cal R}$:
\bea
{\cal R}_{\lambda;pn} ^{ph} = \frac{1}{\omega_{\lambda} - \ve_{p} + \ve_{n}}\sum\limits_{{p}^{\prime}{n}^{\prime}}{\tilde V}_{(-)pn^{\prime},np^{\prime}}^{ph,hp}{\cal R}_{\lambda;p^{\prime}n^{\prime}}^{ph} \nonumber \\
{\cal R}_{\lambda;np} ^{ph} = \frac{1}{\omega_{\lambda} - \ve_{n} + \ve_{p}}\sum\limits_{{p}^{\prime}{n}^{\prime}}{\tilde V}_{(+)np^{\prime}, pn^{\prime}}^{ph,hp}{\cal R}_{\lambda;n^{\prime} p^{\prime}}^{ph},
\label{trden}
\eea
where the upper indices indicate the location of the proton and neutron states with respect to the Fermi energy (FE): 'p' (particle) is above the FE ($\eta$ = 1) and 'h' (hole) is below the FE ($\eta$ = -1). The quantities $\ve_{p,n}$ are the mean-field energies for protons and neutrons, respectively, and the interaction in the isovector channel $\tilde V_{(\pm)}$ is represented by the exchange of positively/negatively charged $\pi$ and $\rho$ mesons, respectively:  
\begin{eqnarray}
{\tilde V}(1,2) =
g_{\rho}^2{\vec\tau}_1{\vec\tau}_2(\beta\gamma^{\mu})_1(\beta\gamma_{\mu})_2
D_{\rho}({\bf r}_1,{\bf r}_2) -\nonumber\\
- \Bigl(\frac{f_{\bf\pi}}{m_{\pi}}\Bigr)^{2}{\vec\tau}_1{\vec\tau}_2({\bf\Sigma}_1{\bf\nabla}_1)
({\bf\Sigma}_2{\bf\nabla}_2)D_{\pi}({\bf r}_1,{\bf r}_2) + \nonumber \\
+
g^{\prime}\Bigl(\frac{f_{\pi}}{m_{\pi}}\Bigr)^2{\vec\tau}_1{\vec\tau}_2{\bf\Sigma}_1{\bf\Sigma}_2
\delta({\bf r}_1 - {\bf r}_2).
\label{tildev}
\end{eqnarray}
After the solution of the Eq. (\ref{trden}), the isovector phonon vertices are obtained as follows:
\bea
\zeta_{\lambda;pn}^{\eta_p\eta_n} =  \sum\limits_{p^{\prime},n^{\prime}} {\tilde V}_{(-)pn^{\prime},np^{\prime}}^{\eta_p h,\eta_n p}{\cal R}_{\lambda;p^{\prime}n^{\prime}}^{ph}\nonumber\\
\zeta_{\lambda;np}^{\eta_n\eta_p} =  \sum\limits_{p^{\prime},n^{\prime}} {\tilde V}_{(+)np^{\prime},pn^{\prime}}^{\eta_n h,\eta_p p}{\cal R}_{\lambda;n^{\prime}p^{\prime}}^{ph},
\label{t-vertices}
\eea
and then their reduced matrix elements and frequencies obtained from Eq. (\ref{trden}) are used in the self-energies (\ref{se1a}), (\ref{se1b}).  Their diagrammatic representation is given in the Fig.  \ref{figse}.  
For both proton and neutron self-energies, the first two terms on the right-hand side represent the contributions from isoscalar vibrations (incoming, outgoing and intermediate single-particle states are of the same isospin). The last two terms are the contributions from isospin-flip vibrations (the intermediate states have different isospin). 

\section{Single-particle states in $^{100,132}$Sn: Details of calculations, results and discussion}

The calculations are performed for $^{100,132}$Sn and the procedure is divided into the three steps. (i) RRPA \cite{RRPA} and pn-RRPA (\ref{trden}) calculations with the parameter set NL3* \cite{NL3*}, with the interaction of Eq. (\ref{tildev}) and $g^{\prime} = 0.6$ are done for determining the frequencies and the reduced transition probabilities of the phonon modes. The modes below 15 MeV for $^{132}$Sn and below 20 MeV for $^{100}$Sn (with respect to the daughter nuclei in the isospin-flip case) with $J^{\pi}_{\mu} = 2^+, 3^-, 4^+, 5^-, 6^+$ in the isoscalar channel (only natural parities are included in the phonon space because spin-flip phonons are known to contribute very little \cite{IBV.12,GRB.05}) and with $J_{\lambda}^{\pi} = 0^{\pm},1^{\pm},2^{\pm}, 3^{\pm}, 4^{\pm}, 5^{\pm}, 6^{\pm}$ in the isovector channel are included in the model space. After that the phonons are selected according to the usual truncation scheme, keeping the modes with the transition probabilities larger than 5\% of the one with the maximal probability. The transition probabilities of the spin-isospin-flip phonons (unnatural parity) are evaluated for the operators of Eq. (\ref{extfield}) with $L = \lambda - 1 $ for $\lambda > 0$ and with $L = \lambda + 1 $ for $\lambda = 0$. For the modes with natural parity the usual multipole operators are assumed.
 (ii) The complementary vertex matrix elements were calculated according to Eq. (\ref{t-vertices}) for the selected isospin-flip phonons, and by a similar procedure \cite{LR.06} for the isoscalar ones. (iii) The obtained frequencies and vertices are included in the neutron and proton self-energies (\ref{se1a}), (\ref{se1b}) and the Dyson equation (\ref{dyson}) is solved in the diagonal approximation. It is transformed to a diagonalization problem,  by the method described in Ref. \cite{RW.73}. The first illustrative calculations presented below neglect the terms in the self-energies (\ref{se1a}), (\ref{se1b}) with $\eta_{p^{\prime}}\neq\eta_n$, $\eta_{n^{\prime}}\neq\eta_p$, which correspond to the backward-going diagrams and contain effects of the ground states correlations (GSC).
These GSC associated with the isovector phonons are represented by the last terms in the diagrammatic form of the self-energies $\Sigma^{(e)}_{pp^{\prime}}$ and $\Sigma^{(e)}_{nn^{\prime}}$ in Fig. \ref{figse}. The GSC-terms associated with the isoscalar phonons and given by the second graphs on the r.h.s. of  Fig. \ref{figse} are included in the calculations. 

\begin{figure}
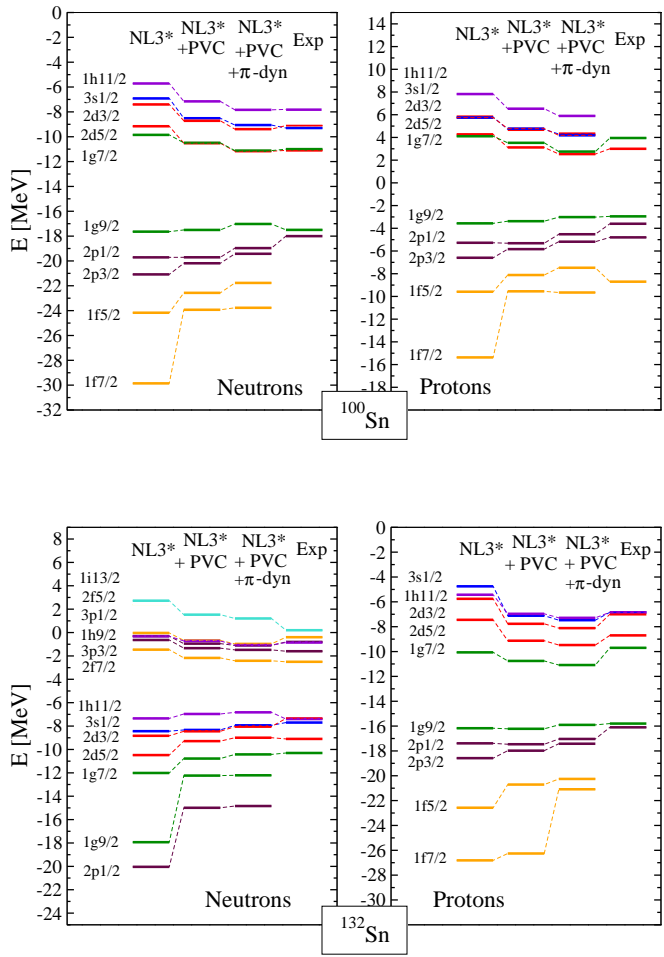

\begin{center}

\vspace{0.5cm}

\includegraphics[scale=0.35]{100sn_pion_nogsc_4.eps}

\vspace{1cm}

\includegraphics[scale=0.35]{132sn_pion_nogsc_4.eps}
\end{center}
\caption{Single-particle states in $^{100,132}$Sn calculated in the RMF and in extended approaches (see text for details).}
\label{fig3}%
\end{figure}
The results of the calculations for the single-particle states in $^{100,132}$Sn are displayed in Fig. \ref{fig3}. The left columns (NL3*) in each panel correspond to the mean-field case, the second left columns (NL3*+PVC) show the RPVC calculations  including only isoscalar phonons, the second right columns (NL3*+PVC+$\pi$-dyn) contain the results obtained with the additional contribution of the isovector phonons, and 'experimental' single-particle energies  extrapolated from data \cite{GBST.14,GLM.07} are presented in the right columns (Exp). Notice that the two middle columns show only the dominant single-particle states, i.e. those with the maximal spectroscopic amplitudes.  

The influence of the isoscalar phonons on the single-particle spectra was investigated systematically within the RPVC model in Ref. \cite{LA.11} and a significant overall improvement of the description of the dominant single-particle states was demonstrated. The columns (NL3*+PVC) correspond to this approach and are to be compared to the next columns (NL3*+PVC+$\pi$-dyn) in order to reveal the dynamical contribution of pions to the positions of the dominant single-particle states. The contributions of the isospin-flip phonons are associated with mainly pionic processes, because the analysis of the matrix elements of the interaction (${\tilde V}_\rho + {\tilde V}_\pi + {\tilde V}_{\delta \pi}$) in Ref. \cite{LZRRM.12} shows that the contribution from the $\rho$-meson exchange is relatively small. 

Thus, as can be seen from Fig. \ref{fig3}, the contribution of the pion dynamics provides additional shifts of the dominant single-particle states, which ranges from a few hundred keV to 1 MeV. The spectroscopic factors for the majority of the considered states change only little compared to those of Ref. \cite{LA.11}. However, for the states far from the Fermi surface like, for instance, the proton state 1f$_{7/2}$ in $^{132}$Sn, the effect can look really dramatic, because of a redistribution of the strength and a change of the dominant fragment. 

It follows from Eqs. (\ref{se1a},\ref{se1b}) that the effect of various contributions to the nucleonic self-energies is determined by the matrix elements 
$\gamma_{(\mu;nn^{\prime})}^{\eta_n\eta_{n^{\prime}}}$, $\gamma_{(\mu;pp^{\prime})}^{\eta_p\eta_{p^{\prime}}}$ and $\zeta_{(\lambda;np^{\prime})}^{\eta_n\eta_{p^{\prime}}}$, $\zeta_{(\lambda;pn^{\prime})}^{\eta_p\eta_{n^{\prime}}}$  given by Eq. (\ref{gammazeta}),  and by the energy denominators in Eqs. (\ref{se1a},\ref{se1b}), which include combinations of single-particle mean-field energies $\varepsilon_{p^{\prime}}, \varepsilon_{n^{\prime}}$ and phonon frequencies: $\Omega_{\mu}$ for the isoscalar phonons and $\omega_{\lambda}$ for the isovector ones. Although the self-energies of Eqs. (\ref{se1a},\ref{se1b}) do not directly provide the splittings and shifts of the single-particle states and the Dyson equation (\ref{dyson}) has to be solved for this, in the second-order perturbation theory the matrix elements of $\Sigma^{(e)}_k$ can be estimated by setting $\varepsilon = \varepsilon_k$ in Eqs. (\ref{se1a},\ref{se1b}), where the index $k$ stands for both neutron and proton single-particle states. Then it is clear, that the largest contributions are given by the terms with the largest phonon vertices and the lowest phonon frequencies, that determines the above mentioned truncation. The typical values of the $\gamma_{\mu}$ are between 0.1 MeV and 1 MeV, in rare cases achieving few MeV. For the vertices $\zeta_{\lambda}$ the pn-RRPA gives somewhat smaller (factor 2-3) values in average, while in both isoscalar and isovector cases the largest matrix elements are obtained for the lowest multipoles. The existence of the isoscalar collective phonons with frequencies below 10 MeV is the most obvious reason for the strong impact of PVC on the single-particle spectra (see numerous papers cited in the introduction). In Fig. \ref{fig3} this effect on the dominant states can be seen by comparison of the first left and the second left columns of each panel. The typical frequencies of isovector spin-flip phonons of the lowest multipoles are seen in Fig. \ref{fig1}  which shows that they can be quite low (in particular, 2$^-$ states shown by green curves). The second-order self-energy denominators corresponding to isoscalar and isovector phonons have values of the same order of magnitude. In the latter case the phonon frequencies $\omega_{\lambda}$ add up to proton-neutron single-particle energy differences $\Delta_{pn^{\prime}} = \varepsilon_p - \varepsilon_{n^{\prime}}$, $\Delta_{np^{\prime}} = \varepsilon_n - \varepsilon_{p^{\prime}}$, producing the energy denominators 
$D_{pn^{\prime}\lambda} = \Delta_{pn^{\prime}}  \mp \omega_{\lambda}$, $D_{np^{\prime}\lambda} = \Delta_{np^{\prime}}  \mp \omega_{\lambda}$. Among many terms with large denominators of several tens MeV, there are usually a few of them with the energies below 10 MeV,  which give the leading contributions for both IS and IV phonons.
\begin{figure}
\begin{center}
\vspace{-0.5cm}
\includegraphics[scale=0.47]{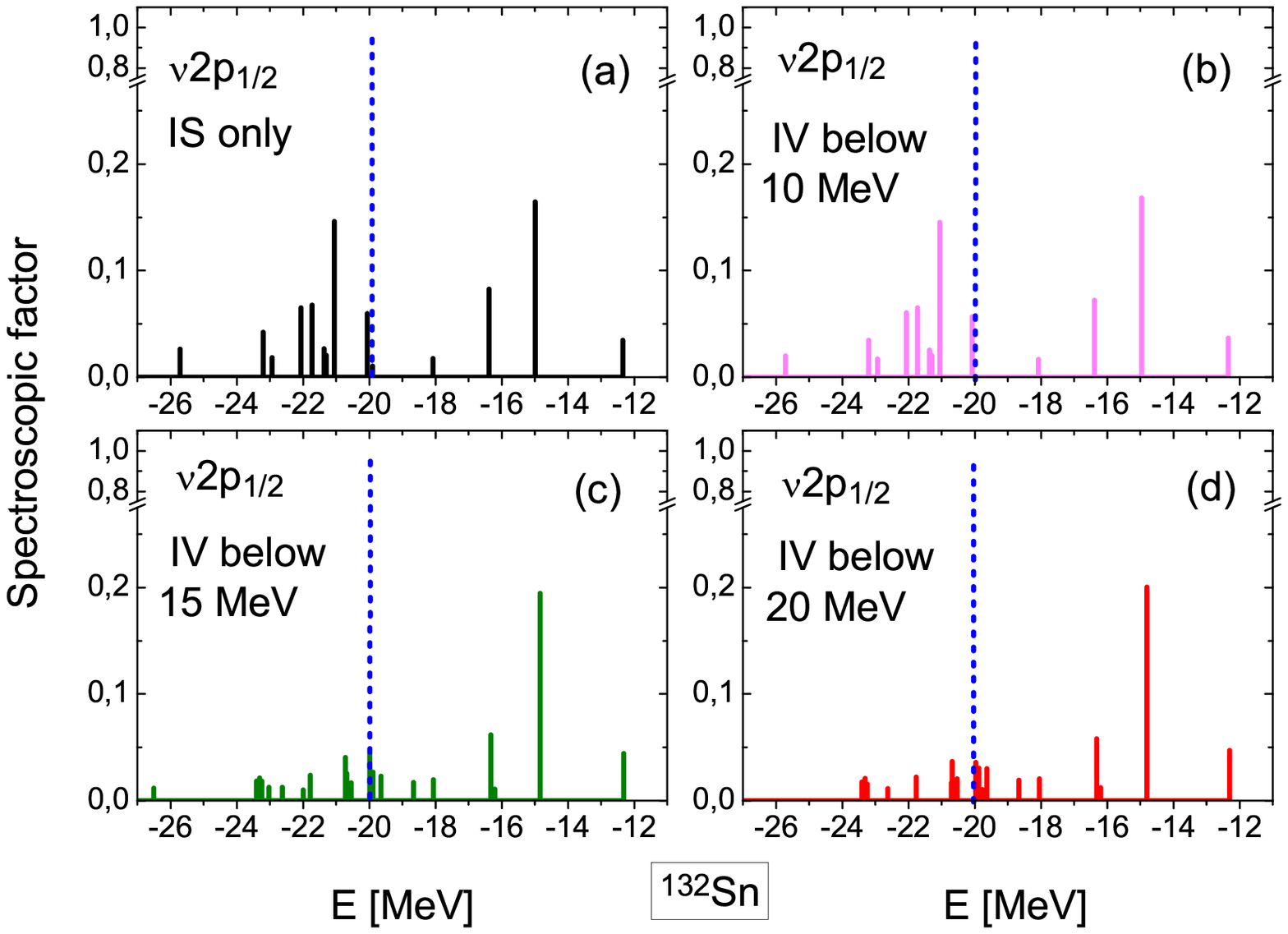}
\vspace{-1cm}
\end{center}
\caption{Strength distributions of the neutron 2p$_{1/2}$ deep hole state in $^{132}$Sn calculated with truncations with respect to the frequencies of the isovector phonons, compared to the RMF strength (dashed blue).}
\label{fig4}%
\end{figure}
\begin{figure}
\begin{center}
\vspace{-0.5cm}
\includegraphics[scale=0.47]{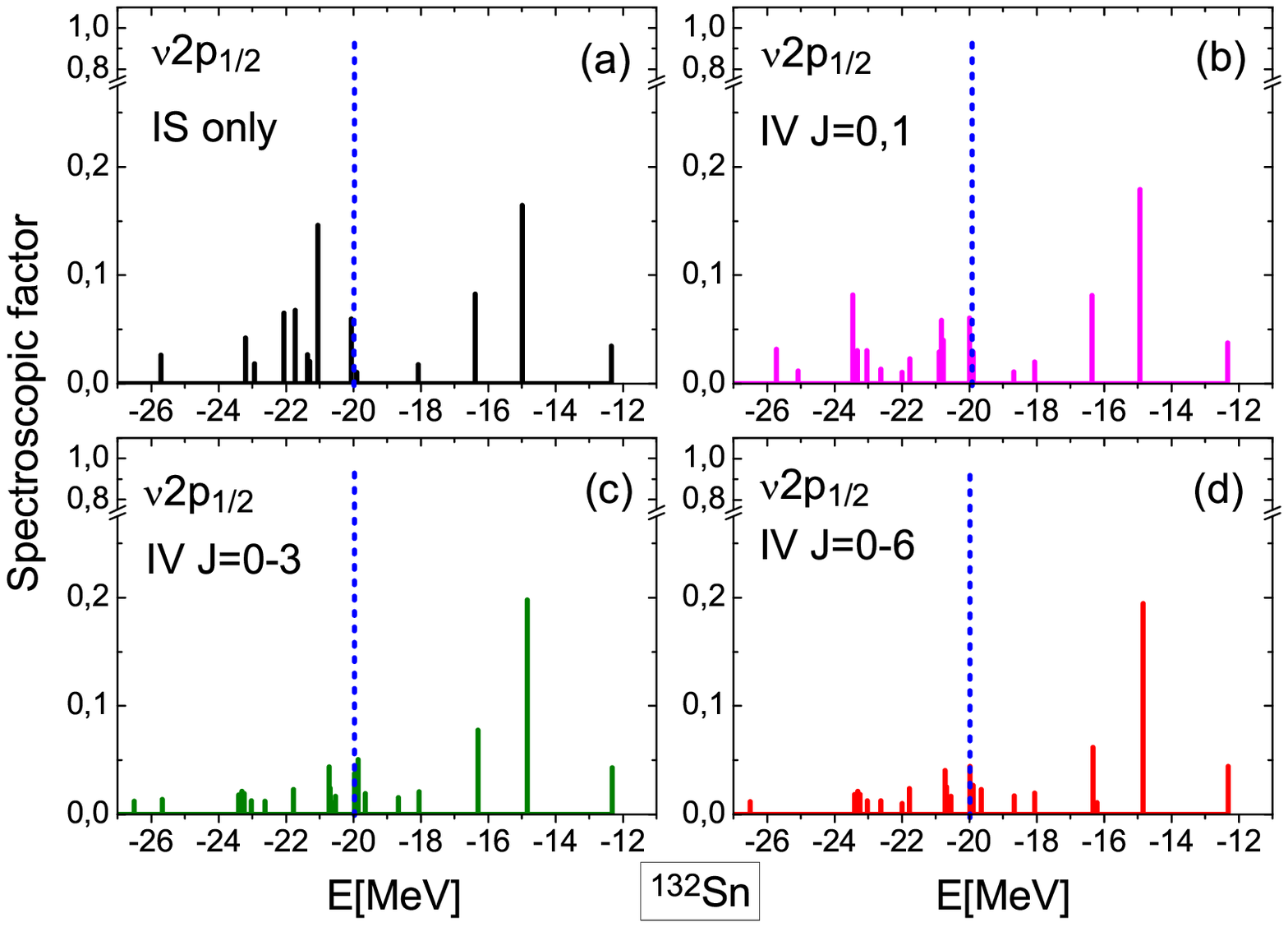}
\vspace{-1cm}
\end{center}
\caption{Strength distributions of the neutron 2p$_{1/2}$ deep hole state in $^{132}$Sn calculated with truncations with respect to the multipolarities of the isovector phonons, compared to the RMF strength (dashed blue). }
\label{fig5}%
\end{figure}
\begin{figure}
\begin{center}
\vspace{-0.5cm}
\includegraphics[scale=0.47]{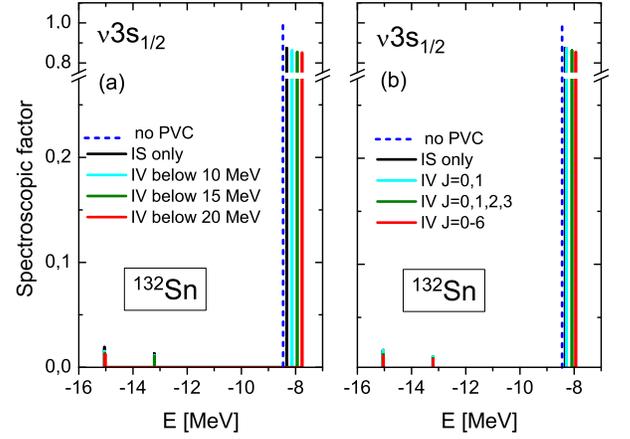}
\vspace{-1cm}
\end{center}
\caption{Strength distributions of the neutron 3s$_{1/2}$ hole state in $^{132}$Sn calculated with various truncations, compared to the RMF strength (dashed blue). Panel (a): truncation by energy (multipoles $J_{\lambda}$ = 0-6 included); panel (b): truncation by $J_{\lambda}$ (energies up to 15 MeV included).}
\label{fig6}%
\end{figure}

Figures \ref{fig4}, \ref{fig5}, \ref{fig6} illustrate a quantitative assessment of the contributions of isovector phonons with various multipolarities and frequencies to the single-particle characteristics. Those figures show single-particle strength distributions for the two neutron states of $^{132}$Sn: 2p$_{1/2}$ (Figs. \ref{fig4}, \ref{fig5}) and 3s$_{1/2}$ (Fig. \ref{fig6}). The latter represents a so-called "good single-particle state" because of its closeness to the Fermi edge, and the former is an example of a "deep hole state", which is far from it. 

Fig. \ref{fig4} shows how the strength distribution of the 2p$_{1/2}$ state evolves with the energy cut-off of the phonon space, when: only isoscalar (IS) phonons are included in the model space (a); isovector (IV) phonons below 10 MeV (b), 15 MeV (c) and 20 MeV (d) with $J_{\lambda} = 1-6$ are included in addition to the IS ones, in comparison to the initial mean-field state (dashed blue). One can see that the 2p$_{1/2}$ state is already fragmented considerably by coupling to IS phonons, and two major fragments can be distinguished at -15.0 MeV and at -21.0 MeV. The inclusion of the lowest IV phonons below 10 MeV affects the distribution very little, but with adding IV phonons of frequencies between 10 and 15 MeV one can see further fragmentation and disappearance of the second major fragment at -21.0 MeV, while the first fragment at -15.0 MeV remains almost unchanged and finally represents the dominant 2p$_{1/2}$ state. This strength distribution saturates at 15 MeV truncation: the distributions 4(c) and 4(d) are almost identical. 

Fig. \ref{fig5} displays the strength distributions of the same 2p$_{1/2}$ neutron state for the cases when only IS phonons are included in the model space (a) and when IV phonons with $J_{\lambda}^{\pi} = 0^{\pm}, 1^{\pm}$ (b), $J_{\lambda}^{\pi} = 0^{\pm}, 1^{\pm}, 2^{\pm}, 3^{\pm}$ (c) and $J_{\lambda}^{\pi} = 0^{\pm} - 6^{\pm}$ (d) are included up to 15 MeV, compared to the initial mean-field state (dashed blue). These figures illustrate the role of various multipoles of the IV phonons in the single-particle strength distribution. Similarly to the case of IS phonons, known from literature, low-J modes give the major contribution to the strength redistribution while the phonons with $J_{\lambda}^{\pi} = 4^{\pm}, 5^{\pm}, 6^{\pm}$ produce a weaker effect, that justifies the truncation of the phonon basis at $J_{\lambda}=6$.

The change of the 3s$_{1/2}$ state because of the IV-type PVC, which is shown in Fig. \ref{fig6}, reduces to a gradual move toward the Fermi surface with every extension of the phonon space and a very little fragmentation. The effect also becomes smaller when the phonon energy increases while no clear saturation is seen for this state for the IV phonons up to 20 MeV. This points out that a proper regularization may be needed for the considered mass operator, for instance, a subtraction technique analogous to the one employed in the response theory \cite{LRT.08} can be adopted.  

Another consequence coming from the analysis above is that there is no clear dominance of the lowest IV phonons. This emphasizes the distinction between the energies of the most important IS modes and those of the of IV modes, due to the fact that the relevant energy  differences  between neutron or proton states are typically smaller  than those between proton-neutron states.

In conclusion, except for a few cases, the shifts of the dominant states in Fig. \ref{fig3} are toward the Fermi surface and give an additional improvement of the description, compared to available data.  Proton particle states in $^{132}$Sn seem to be very sensitive to the PVC effects, which tend to overestimate the shifts, however, the obtained picture can change further after the inclusion of ground state correlations associated with the IV phonons, which will be addressed by future efforts. 

\section{Summary and outlook }

The dynamical contribution of the pion is included into the single-particle self-energy of a finite nuclear system in a self-consistent relativistic framework. It has been shown that in medium-mass nuclei low-lying isospin-flip states associated with soft pionic modes can occur with sizable transition probabilities and, therefore, their coupling to single-particle and other collective modes should be considered. The strength of this coupling and its effect on the single-particle states is evaluated for $^{100,132}$Sn and the shell structure of these nuclei is found to be sensitive to pion dynamics. Thereby, (i) pion-nucleon correlations beyond the Hartree-Fock approximation are included in the theory based on the QHD Lagrangian; (ii) the latter can provide a link between the relativistic nuclear field theory and chiral effective field theories; 
(iii) the relativistic particle-vibration coupling model is extended self-consistently by incorporating isospin-flip phonons. 

As the nucleonic self-energy of the presented approach contains a full summation of the ring diagrams with pion exchange, the pion contribution is included non-perturbatively, although its soft modes entering the nucleonic self-energy are obtained by calculations on the RPA level. 
%
The next natural step will be an inclusion of the ground state correlations associated with the exchange of the isospin-flip phonons, which will provide a more accurate description of the effects considered in the present work. Pairing correlations of the superfluid type should be taken into account for the isovector phonons, which will allow for  more systematic studies of the effects of pion dynamics in open-shell nuclei. The considered effects of pion dynamics should have also an impact on nuclear response in various channels, therefore, a corresponding extension of the response theory can be another exciting future development.
  
\section{Acknowledgement}
Enlightening discussions with E.E. Kolomeitsev, P. Ring, V. Tselyaev, T. Otsuka, and V. Zelevinsky are gratefully acknowledged. The author is very thankful to T. Marketin for providing a part of the code for pn-RRPA matrix elements. This work was supported by US-NSF grants PHY- 1204486 and PHY-1404343.

\end{document}